\ifdefined\WORDCOUNT
\documentclass[superscriptaddress,twocolumn,preprintnumbers,amsmath,amssymb,prl,nofootinbib,groupedaddress]{revtex4-1}
\else
\documentclass[superscriptaddress,twocolumn,showpacs,preprintnumbers,amsmath,amssymb,prl,groupedaddress]{revtex4-1}
\fi
\usepackage{amsmath}  
\usepackage{amsfonts} 
\usepackage{graphicx} 
\usepackage{hyperref}
\usepackage{bbold}
\usepackage{braket}
\usepackage[normalem]{ulem}
\usepackage[table]{xcolor}
\definecolor{darkblue}{RGB}{0,0,149}
\hypersetup{
    colorlinks=true,
    linkcolor=darkblue,
    filecolor=darkblue,
    urlcolor=darkblue,
    citecolor=darkblue,
    linktocpage=true,
}
\usepackage{url}

\begin{document}
\title{Inference of gravitational field superposition from quantum measurements}
 
\author{Chris Overstreet} 
\thanks{Current affiliation:  Department of Physics \& Astronomy, Johns Hopkins University, Baltimore, Maryland 21218, USA}
\affiliation{Department of Physics, Stanford University, Stanford, California 94305, USA}

\author{Flaminia Giacomini}
\affiliation{Perimeter Institute for Theoretical Physics, 31 Caroline St. N, Waterloo, Ontario, N2L 2Y5, Canada}
\affiliation{Institute for Theoretical Physics, ETH Z{\"u}rich, 8093 Z{\"u}rich, Switzerland
}

\author{Joseph Curti} 
\thanks{These authors contributed equally to this work.}
\affiliation{Department of Physics, Stanford University, Stanford, California 94305, USA}

\author{Minjeong Kim}
\thanks{These authors contributed equally to this work.}
\affiliation{Department of Physics, Stanford University, Stanford, California 94305, USA}

\author{Peter Asenbaum} 
\thanks{Current affiliation:  Institute for Quantum Optics and Quantum Information (IQOQI) Vienna, Austrian Academy of Sciences, Boltzmanngasse 3, 1090 Vienna, Austria}
\affiliation{Department of Physics, Stanford University, Stanford, California 94305, USA}

\author{Mark A. Kasevich} 
\email{kasevich@stanford.edu}
\affiliation{Department of Physics, Stanford University, Stanford, California 94305, USA}

\date{\today}

\begin{abstract}
    Experiments are beginning to probe the interaction of quantum particles with gravitational fields beyond the uniform-field regime.  In non-relativistic quantum mechanics, the gravitational field in such experiments can be written as a superposition state.  We empirically demonstrate that semiclassical theories of gravity can avoid gravitational superposition states only by decoupling the gravitational field energy from the quantum particle's time evolution.  Furthermore, such theories must specify a preferred quantum reference frame in which the equations of motion are valid.  To the extent that these properties are theoretically implausible, recent experiments provide indirect evidence that gravity has quantum features.  Proposed experiments with superposed gravitational sources would provide even stronger evidence that gravity is nonclassical.  
\end{abstract}

\maketitle 

\section{I.  Introduction}

Understanding the fundamental nature of gravity is a significant open problem in theoretical physics.   Unlike the other interactions, gravity lacks a satisfactory description as a quantum field, leaving open the possibility that gravity is fundamentally classical \cite{Penrose2014} rather than quantum. On the other hand, several arguments \cite{Bronstein2012,Rickles2011,Zeh2011,Carlip2008,Bose2017,Marletto2017} suggest that gravity ought to have quantum features.  

For many years, it was believed that experiments could not contribute meaningfully to this discussion.  Owing to the weakness of the gravitational interaction, it is not feasible to detect a single graviton in a gravitational-wave observatory or a particle detector \cite{Smolin:1982jt, Smolin:1983ja, Dyson2013}.  Experimental evidence of the quantum nature of gravity is expected to appear at the energy scale corresponding to the Planck mass ($10^{19}$~GeV), which is far beyond the reach of collider experiments.  

Although the quantization of gravity cannot be observed directly, it may be possible to probe gravity's fundamental nature in other ways. For example, two proposals~\cite{Bose2017,Marletto2017} (``BMV experiments'') suggest searching for entanglement generation in a pair of matter-wave interferometers that interact gravitationally. Such low-energy tests might open the first observational window on the quantum description of the gravitational field \cite{Rijavec2021}, but the precise theoretical implication of these experiments is still an open question~\cite{Belenchia2018,Belenchia2019,Christodoulou2019,Hall2018,Anastopoulos2018,Howl2021,Marshman2020,Krisnanda2020,Marletto2020,Galley2020,Pal2021,Kent2021,Zhou2022,Danielson2022,Christodoulou2022,Bose2022,Polino2022,Christodoulou2022a}. The application of one of the fundamental theorems of quantum information, stating that local operations and classical communication cannot generate entanglement \cite{Horodecki2009}, to these experiments leads to the conclusion that the gravitational field cannot be classical if entanglement is observed \cite{footnote1}.  The experiments described in these proposals are challenging to implement because they require large spatial superpositions of large test masses ($\sim 10^{12}$~amu).  For comparison, the largest particle for which matter-wave interference has been demonstrated has a mass of $2.5 \times 10^4$ amu \cite{Fein2019}.  Nevertheless, the possibility of learning about the nature of gravity from such experiments motivates an assessment of what we can infer from experiments accessible with existing technologies.  

Beginning with the COW experiment in 1975 \cite{Colella1975}, many matter-wave interferometers have demonstrated sensitivity to inertial forces.  Until recently, these experiments operated in a regime where the gravitational field is approximately uniform. In any theory respecting the equivalence principle \cite{DiCasola2015}, a uniform gravitational field does not cause relative acceleration between a test particle and a measuring device.  Therefore, the relative accelerations measured in such experiments are caused by the non-gravitational forces that keep the measuring device fixed to the surface of the Earth.  In order to observe genuine gravitational effects in a quantum system, it is necessary to measure a non-uniform gravitational field across the quantum state:  matter-wave interferometers in the uniform-field regime test the equivalence principle \cite{Asenbaum2020a,Bothwell2022} but do not provide quantum tests of any other gravitational properties \cite{Chryssomalakos2003}.  Throughout this work, we will assume that the equivalence principle is valid to the accuracy of the experimental results we describe.   

Recent experiments have begun to investigate the non-uniform regime of the gravitational field. Observations of phase shifts associated with gravitational tidal forces \cite{Asenbaum2017} and a gravitational Aharonov-Bohm effect \cite{Overstreet2022} indicate that the trajectory \cite{Asenbaum2017} and action \cite{Overstreet2022} of a matter wave in a nontrivial gravitational field are correctly predicted by quantum mechanics. 

Taken by themselves, these results might appear to provide little information about gravity's fundamental nature. However, given the difficulty of performing a more conclusive experiment \cite{Aspelmeyer:2022fgc}, it is useful to assess what can be inferred by combining state-of-the-art experimental results with additional principles about the behavior of gravity. This work can shed light on the fundamental structure of a theory at the interface between quantum theory and gravity while providing a roadmap to prioritize experimental efforts \cite{Marletto2017Nature}.  Here we identify the following three principles:  

\begin{enumerate}
    \item \textit{Existence of gravitational fields}: Any massive particle that is well-localized at a position $\mathbf{x_0}$ sources a gravitational field $\mathbf{g}$ with functional form $\mathbf{g}(\mathbf{x} - \mathbf{x_0})$.
    % \vspace{0.1in}
    \item \textit{Field energy principle}.  The phase of an interferometer is a function of the energies of the fields that interact with the interfering particle.  
    % \vspace{0.1in}
    \item \textit{Quantum relativity principle}.  The laws of physics take the same form in every reference frame, including the reference frames associated with quantum particles (quantum reference frames~\cite{Giacomini2019}).    
\end{enumerate}

Principle (1) asserts that the gravitational interaction is mediated by a field but makes no assertion about the gravitational field sourced by a particle that is not well-localized.  

The quantum relativity principle (QRP) encodes the relational nature of physical statements. This means that all physical quantities are expressed in terms of the relationship between physical systems, without appealing to an absolute background structure. 
In the case of two particles sourcing the gravitational field in the linearized regime, we show in Appendix 1 that the standard description of the gravitational field allows us to describe the interaction as sourced by either one of the two particles, with the other particle moving in this external potential. 
The QRP extends the relativity principle to the case in which the two particles are in a quantum superposition state relative to one another.

This article analyzes the  gravitational Aharonov-Bohm experiment \cite{Overstreet2022} with the goal of determining whether current experimental results, plus the three principles outlined above, are compatible with a classical description of gravity.  We argue that they are not \cite{footnote3}. The article also presents new experimental data that places constraints on certain classes of semiclassical gravitational theories.  First, we show that the quantum-mechanical description of the gravitational Aharonov-Bohm experiment is compatible with all three of the above principles. In the quantum-mechanical picture, the gravitational field in the experiment is described as entangled with the superposed matter degrees of freedom.
Next, we present data demonstrating the insensitivity of the interferometer phase to the probability distribution of the test particle---as predicted by quantum mechanics, but in tension with some semiclassical theories. 
We assess the theoretical reach of the experiment by comparison to future searches for gravitationally mediated entanglement.   
Finally, we argue that in order to deny the existence of a gravitational superposition state in this apparatus, a theory in which gravity is treated as a physical system must either reject principle (1) or reject both principles (2) and (3).

The remainder of the article is organized as follows. Section II presents a technical overview and simplified theoretical model of the gravitational Aharonov-Bohm experiment. Next, we illustrate our principles in the context of quantum mechanics: Section III describes how the gravitational phase shift can be computed from the field energy; Section IV presents new experimental data that are consistent with quantum-mechanical predictions; Section V analyzes the experiment in a quantum reference frame \cite{Giacomini2019} associated with the position of the interfering particle; and Section VI compares the gravitational Aharonov-Bohm experiment to the proposed BMV experiments. Finally, Section VII discusses how our principles (1)-(3) lead to a superposition of gravitational fields and explains how these assumptions are violated in alternative theories of gravity.   

\section{II.  Experimental overview and simplified theoretical model}

The idea of the gravitational Aharonov-Bohm experiment \cite{Overstreet2022} is to observe the gravitational interaction between a quantum test particle and a classical source mass in the regime where the wave packet separation of the test particle is larger than the distance to the source mass.  In this regime, the phase response of the interfering test particle becomes decoupled from the forces on the interferometer arms.  In particular, one can observe a gravitational phase shift even in a configuration where the deflection-induced phase contribution vanishes, as demonstrated experimentally by independent measurements of the arm deflections \cite{Overstreet2022}.  For the purpose of this work, the significance of the experimental result is that it is consistent with the predictions of standard quantum mechanics for the behavior of a quantum particle in a nontrivial gravitational field.   

\begin{figure}[h]
\centering
\includegraphics[width=1.0\linewidth]{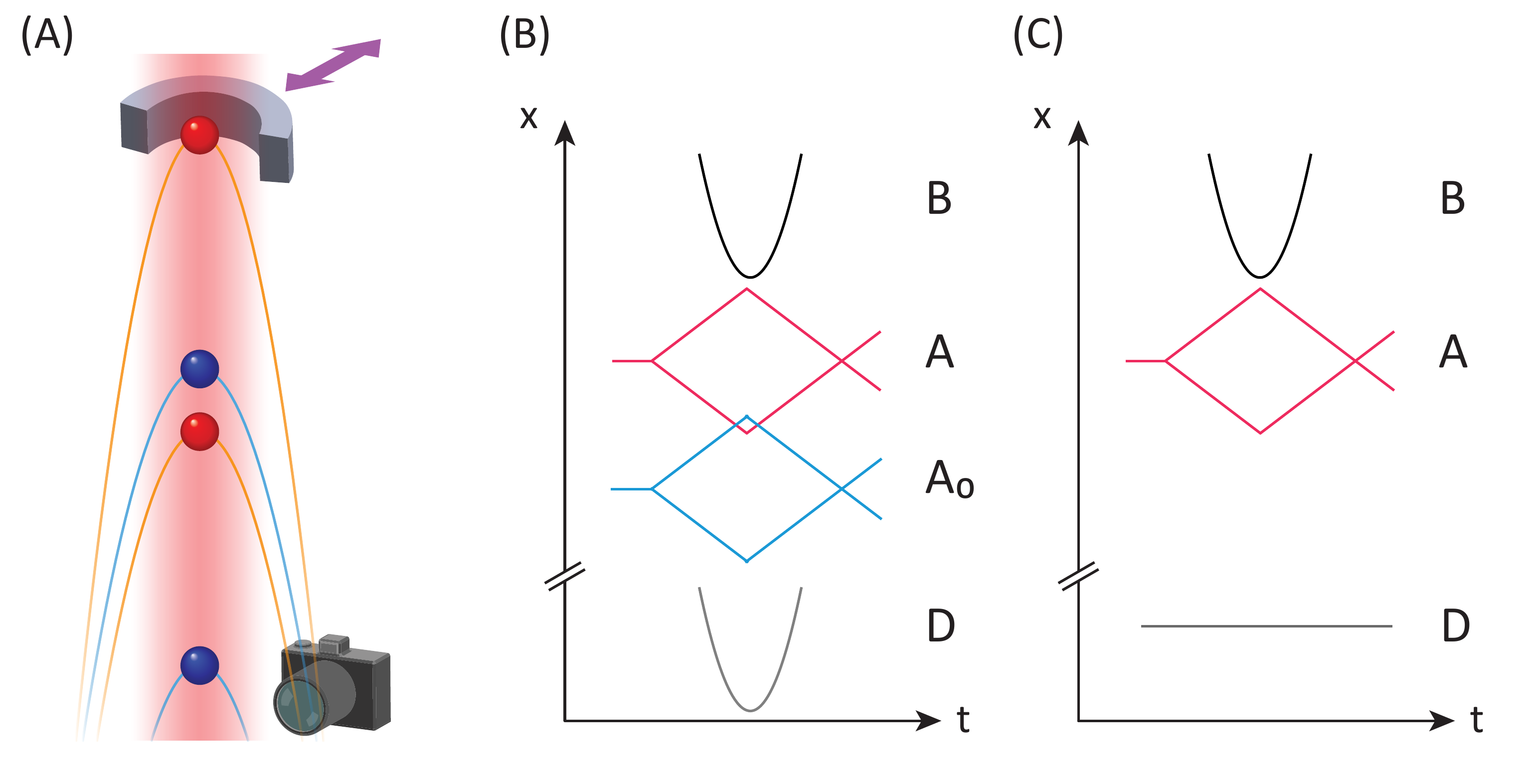}
\label{Fig:1}
\caption{(A):  Experimental schematic.  The phase shift of a matter-wave gradiometer due to the gravitational interaction with a source mass is measured by detecting output port population ratios with and without the source mass installed. The atomic trajectories are horizontally exaggerated for clarity. (B):  Spacetime trajectories of test particles A and A$_0$, source mass B, and detector D in a freely falling reference frame.  Interferometer A is used to probe the gravitational field of source mass B.  A second interferometer A$_0$ is used to suppress technical noise sources.  (C) Simplified model of experiment, neglecting the lower interferometer and non-gravitational acceleration of detector. Figures are not to scale.}
\end{figure}

In the experiment \cite{Overstreet2022} (Fig. 1A), an ultracold cloud of $^{87}$Rb is used as the input of a single-source gradiometer \cite{Asenbaum2017} with baseline $24$ cm and wave packet separation $25$ cm.  The upper interferometer in the gradiometer is sensitive to the gravitational interaction with the tungsten source mass, while the lower interferometer mainly acts as a phase reference.  The matter-wave beam splitters and mirrors are implemented by applying a sequence of laser pulses to the atoms, inducing sequential $2\hbar k$ Bragg transitions.  The population of each output port is detected by fluorescence imaging.  Fig. 1B shows the spacetime trajectories of the source mass, interferometers, and detector in a freely falling reference frame.  For a full description of the experimental apparatus and measurement protocol, see Ref.~\cite{Overstreet2022}.

In this section, we give a standard description of the gravitational Aharonov-Bohm experiment using a quantum-mechanical model that implements several simplifying assumptions.  First, we ignore the back action of the test particles on the source mass trajectory, and we also ignore any perturbations of the source mass induced by its interaction with the environment.  Second, we neglect the gravitational interaction between the source mass and the lower interferometer, allowing us to exclude the lower interferometer and the optical phase reference from the model.  Third, we ignore the non-gravitational acceleration of the detector, which does not affect the measurement outcomes.  

Fig. 1C shows the spacetime trajectories of the test mass, source mass, and detector in this model.  A test particle of mass $m$ (``particle A''), initialized in a Gaussian state \cite{footnote4}, is used as the input to a Mach-Zehnder interferometer.  The particle is placed in a superposition of momentum states by a beam splitter at time $t = 0$, which are reflected by a mirror at time $t = T$ and interfered by a second beam splitter at time $t = 2T$.  In the freely falling frame of the atoms, the central trajectories of the two interferometer arms are given by 
\begin{align} \label{eq:trajAfromC}
    x_1(t) &= \begin{cases} x_0 + \frac{\hbar k}{m} t,  & 0\le t \le T\\
    x_0 - \frac{\hbar k}{m} (t - 2T), & T< t \le 2T \nonumber
    \end{cases} \\
    x_2(t) &= \begin{cases} x_0 - \frac{\hbar k}{m} t,  & 0\le t \le T\\
    x_0 + \frac{\hbar k}{m} (t - 2T), & T< t \le 2T 
    \end{cases}, 
\end{align}
where $\hbar$ is the reduced Planck's constant and $k$ is the magnitude of the beam splitter wave vector.  The details of the beam splitter and mirror implementation are unimportant for our analysis.  

During the interferometer, a source particle of mass $M$ (``particle B''), initialized in a Gaussian state, is brought close to one interferometer arm and then removed.  For concreteness, we will assume that the source particle follows a parabolic trajectory: 
\begin{equation} \label{eq:trajBfromC}
    x_s(t) = \frac{1}{2} a (t - T)^2 + x_{s,0}.    
\end{equation}
A third particle (``particle D'') with mass $M_D$ defines the reference frame.  We assume that particle D is far enough away from particles A and B that its gravitational influence is negligible.  

In quantum mechanics, and in accordance with our principle (1), particles A and B interact gravitationally with potential energy 
\begin{equation}
    U(|\hat{x}_A - \hat{x}_B|) = -\frac{G m M}{|\hat{x}_A - \hat{x}_B|},
\end{equation}
where $G$ is Newton's gravitational constant and $\hat{x}_i$ is the position operator of particle $i$.  This expression follows naturally from combining quantum field theory with general relativity \cite{Anastopoulos2014} but is modified in alternative gravitational theories, e.g. in semiclassical gravitational theories.  In the time intervals $(0, T)$ and $(T, 2T)$, the system can be described by the Hamiltonian 
\begin{equation} \label{eq:HamC}
    \hat{H}_{AB}^{(D)} = \frac{\hat{p}^2_A}{2 m} + \frac{\hat{p}^2_B}{2 M} + U(|\hat{x}_A - \hat{x}_B|) - M a \hat{x}_B.
\end{equation}

After the final beamsplitter, the two output ports spatially separate due to their momentum difference.  When the measurement occurs, the two ports are displaced by $d_1$ and $d_2$ from the detector, respectively.  The interferometer phase $\phi$ is defined by the probabilities that the test particle will be observed at distance $d_1$ or $d_2$ from the detector: 
\begin{equation} \label{Eq:intPhase}
    \phi \equiv \arccos\left(\frac{P(d_1) - P(d_2)}{P(d_1) + P(d_2)}\right)
\end{equation}
where $P(d_i)$ is the probability of observing particle A at distance $d_i$ from particle D.  Note that the interferometer phase is defined entirely in terms of relative coordinates, as is necessary for an observable quantity.  
One can compare the interferometer phase with and without the source mass present to measure the phase shift $\Delta \phi$ due to the source mass.  For the purpose of calculating $\Delta \phi$, we can neglect the trajectory perturbations of particle A induced by $U$ \cite{footnote11}. 

The phase shift due to the gravitational interaction between A and B can be computed \cite{Storey1994} as
\begin{align}
    \Delta \phi &= \frac{1}{\hbar} \int_{t = 0}^{2T} \left[ U(|x_1 - x_s|) - U(|x_2 - x_s|) \right] dt  \\
    &= -\frac{G m M}{\hbar} \int_{t = 0}^{2T} \left[ \frac{1}{|x_1 - x_s|} - \frac{1}{|x_2 - x_s|} \right] dt.
\end{align}
In the reference frame of particle D, this phase shift can be understood to arise from the different paths of particle A in the background classical gravitational field sourced by particle B. This means that the phase shift can be rewritten as
\begin{equation}
    \Delta \phi = \frac{m}{\hbar} \left[ \mathcal{V}(x_1) - \mathcal{V}(x_2) \right], \label{Eq:deltaPhiD}
\end{equation}
where $\mathcal{V} (x_i)$ is the classical gravitational potential associated with particle B integrated over each trajectory $x_i$ of particle A in the interferometer.

During the interferometer, the quantum state (in the reference frame of particle D) takes the form
\begin{equation} \label{eq:gravitystateD}
    \ket{\psi}^{(D)}(t) = \frac{1}{\sqrt{2}} \left(\ket{x_1}_A \ket{\alpha_1}_\text{G} + e^{i \theta(t)}\ket{x_2}_A \ket{\alpha_2}_\text{G} \right) \ket{x_s}_B   
\end{equation}
for some $\theta(t)$.  Here the subscript $A$ indicates the state of particle A, and we have written the state of the gravitational field $\ket{\alpha_i}_\text{G}$, where $\alpha_i$ represents the field sourced by point particles of mass $m$ and $M$ at positions $x_i$ and $x_s$, respectively. 
Although the gravitational field depends on the source mass positions, the gravitational field still has a quantum state and an associated Hilbert space.  The necessity of including the gravitational field in the quantum state description is discussed in \cite{Belenchia2018, Belenchia2019, Danielson2022}, and the explicit form of the gravity state is given in \cite{Chen2022}. An analogous earlier study~\cite{Barnich:2010bu} reports the same conclusion for the electromagnetic field.  

In particular, the calculation of the quantum state of gravity for the gravitational AB experiment is analogous to the one of Ref.~\cite{Chen2022}, where the quantum state of gravity was derived for a static source in a quantum superposition of localized states. There, the Newton potential corresponds to the ground state of the free gravitational field Hamiltonian with a matter source, shifted by a phase proportional to the solution of the Gauss equation for a source in the corresponding classical position. For such states of the source, the quantum states of the gravitational field associated with different positions of the source are orthogonal. The Newtonian phase is obtained by evolving the quantum state with this Hamiltonian.

Note that the gravitational field (which is sourced by both particles A and B) is entangled with the position of particle A and is in a spatial superposition with length scale $\hbar k T/m$.  In a general-relativistic context, this state is described by writing a superposition of metrics \cite{Christodoulou2019}, but the Newtonian limit is sufficient for our purposes. 

\section{III.  Phase shift from field energy}

In this section, we show that in quantum mechanics, an interferometer's phase shift can be computed from the energies of the fields that interact with the interfering particle.  First, we consider the electromagnetic case.
The electrostatic interaction energy $U_\text{EM}$ of a particle with charge $q$ in an electric potential $V_\text{EM}$ is given by 
\begin{equation}
U_{\text{EM}} = q V_\text{EM}.
\end{equation}
From a field-theoretic perspective, it is well known \cite{Aharonov1961,Landau1971} that this quantity is equivalent to the electric field energy $E_\text{EM}$, given by 
\begin{equation}
    E_\text{EM} = \frac{1}{2}\epsilon_0 \int |\mathbf{E}|^2 dV \label{Eq:fieldEnergy}
\end{equation}
where $\mathbf{E}$ is the electric field and the integral is taken over all space, with the divergent self-energy of the particle's electric field subtracted from the integrand. Notice that if the electrostatic interaction were due to a non-local potential, and not to a field, it would not be possible to associate the field energy $E_\text{EM}$ with it.

An analogous relationship holds for the classical gravitational field.  In the weak-gravity, non-relativistic limit, the interaction energy $U$ of a particle of mass $m$ in a gravitational potential $V$ is given by 
\begin{equation}
U = m V.
\end{equation}
This interaction energy is equivalent \cite{Landau1971} to the gravitational field energy $E$, which is given by 
\begin{equation}
    E = -\frac{1}{8\pi G} \int |\mathbf{g}|^2 dV \label{Eq:gravFieldEnergy}
\end{equation}
where $G$ is Newton's constant, $\mathbf{g}$ is the gravitational field, and the integral is taken over all space.  As in the electromagnetic case, the integrand of Eq.~\eqref{Eq:gravFieldEnergy} contains contributions from the gravitational field of the test mass as well as the source mass, and the gravitational self-energies have been subtracted out of the integrand.  This definition of the gravitational field energy can be generalized to linearized gravity, provided that one uses a coordinate system that asymptotically approaches Minkowski coordinates \cite{Landau1971}.

The usual approach in quantum mechanics is to account for the gravitational interaction energy by including the potential $V$ in the Hamiltonian.  Alternatively, we can take the perspective that the energy resides in the gravitational field. This amounts to replacing $V$ with $E$ in the Hamiltonian, where $|\mathbf{g}|^2$ is promoted to an operator that acts on the gravitational field's quantum state.  

As discussed in Section 2, the interferometer phase shift of a particle of mass $m$ evolving in a perturbing gravitational potential $V$ is approximately given by  
\begin{equation}
    \Delta \phi = \frac{m}{\hbar} \int \left[ V(x_1) - V(x_2) \right] dt
\end{equation}
where $x_1(t)$ and $x_2(t)$ are the trajectories of the interferometer arms.  Equivalently, the phase shift can be written as
\begin{equation}
    \Delta \phi = \frac{1}{\hbar} \int (E_1 - E_2)\, dt \label{Eq:intFieldEnergy}
\end{equation}
where $E_i(t)$ is $E$ evaluated on the quantum state of the gravitational field associated with trajectory $x_i(t)$.  Quantum mechanics thus satisfies the field energy principle.  

Crucially, the gravitational field energy cannot give rise to the experimentally observed relative phase in this way unless the gravitational field is in a superposition state that is entangled with the trajectory of the interfering particle.  (If instead the gravitational field is in a classical state, then the gravitational field is not entangled with the interfering particle's trajectory, so $E_1 = E_2$ and $\Delta \phi = 0$.)
% For a classical gravitational field, in Eq.~\eqref{Eq:intFieldEnergy}, we would have $E_1 = E_2$ and $\Delta \phi = 0$.
Finally, note that the gravitational energy difference between interferometer arms cannot be observed when the gravitational field is approximately uniform at the experimental length scale \cite{Colella1975}.  In that case, the phase shift arises from the relative acceleration between the interfering particles and the phase reference \cite{Wolf2011}, which is induced by non-gravitational forces.

\section{IV.  Independence of phase shift from probability distribution}

We next consider whether the field energy principle can be satisfied in alternative gravitational theories that lack gravitational superposition states. In semiclassical gravitational theories, the gravitational field is represented as a classical state \cite{Carlip2008}. Since the gravitational field in such theories is not entangled with the interferometer trajectories,  Eq.~\eqref{Eq:intFieldEnergy} cannot hold.  It is possible that the interferometer phase shift could still be a function of the gravitational field energy, but this would require modifications of the laws of physics, e.g. a dependence of the phase shift on the probability distribution of the interfering particle.    

\begin{figure}[h]
\centering
\includegraphics[width=1.0\linewidth]{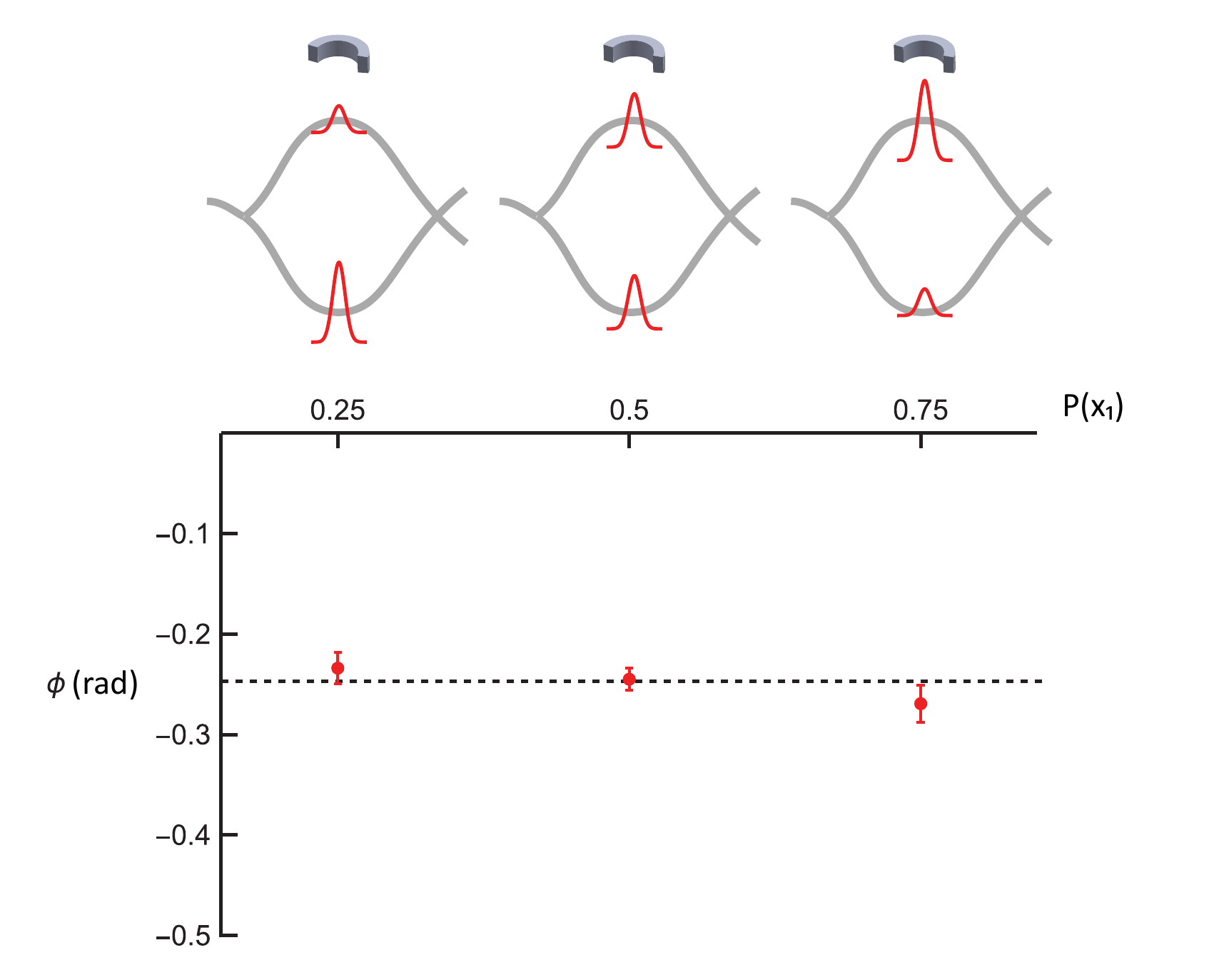}
\caption{\label{Fig:2}Red points:  phase shift induced by tungsten source mass in $52\hbar k$ gradiometer with probability distribution biased toward or away from source mass.  The x axis represents the probability of observing the test particle on the upper interferometer arm, closer to the source mass.  Error bars represent $1\sigma$ statistical uncertainty.  Black dashed line:  quantum-mechanical phase shift prediction.}
\end{figure}

To determine whether a gravitating particle's phase evolution depends on its probability distribution, we measured the phase shift of a $^{87}$Rb $52 \hbar k$ interferometer with $25$ cm wave packet separation.  A second interferometer, located $24$ cm below the first (as illustrated in Figs. 1A and 1B), was used to suppress technical noise sources associated with vibrations and laser phase fluctuations.  The upper arm of the upper interferometer was launched to a height of $4$ cm above a tungsten source mass ($1.25$ kg semicircular ring).  The distance of closest approach to the source mass was $7$ cm.  For a detailed description of the apparatus, see Ref.~\cite{Overstreet2022}.

In this measurement, an asymmetric probability distribution was created by altering the Rabi frequency of the initial beam splitter pulse.  Systematic effects, e.g. from ac-Stark shifts, were suppressed by comparing the gradiometer phase with and without the source mass installed.  Imaging-related systematic errors were further reduced by reversing the direction of the horizontal detection fringe \cite{Asenbaum2020a}.  

Fig.~\ref{Fig:2} shows the phase shift induced by the tungsten source mass as a function of the probability on the upper interferometer arm.  Each point in Fig.~\ref{Fig:2} represents the phase shift between these two configurations.  Within the experimental resolution, there is no statistically significant correlation between the test particle's center of mass and the interferometer phase shift (slope $-0.07 \pm 0.05$, $p = 0.39$).  Moreover, the data are consistent with the quantum-mechanical prediction (reduced $\chi^2 = 1.1$).  
This apparent independence of phase shift from center-of-mass position demonstrates that semiclassical gravitational theories cannot satisfy the field energy principle.  
A specific semiclassical model that is constrained by this result is discussed in Appendix 2.

\section{V.  The relativity of superposition}

In classical mechanics, every particle is associated with a reference frame\,--\,a coordinate system in which that particle is located at the origin.  With appropriate coordinate transformations, any physical interaction can be described in the reference frame of any particle.  This property also holds in quantum mechanics, where the reference frame associated with a quantum particle is called a quantum reference frame (QRF). QRFs are widely expected to be  necessary in a quantum theory of gravity, where the classical, idealized notion of reference frame is no longer sufficient. They have been discussed since 1967 in the context of quantum gravity~\cite{DeWitt:1967yk, Kuchar:1990vy, Brown:1994py, Brown:1995fj, Rovelli:1990pi, Rovelli:2004tv, Rovelli:1995fv, Dittrich:2004cb, Tambornino:2011vg} and quantum information~\cite{Aharonov:1967zza, Aharonov:1984zz, Bartlett:2006tzx, spekkens_resource, Palmer:2013zza, smith_quantumrf, Poulin:2006ryq, busch_relational_1, Loveridge:2016tnh, Loveridge:2017pcv, angelo_1}. Here, we take the formulation introduced in Ref.~\cite{Giacomini2019}, which formalizes the transformation between two QRFs as a quantum superposition of reference frame transformations $\hat{S}$ (see also later related work~\cite{Vanrietvelde:2018pgb, Vanrietvelde:2018dit, Hohn:2018toe, Giacomini:2018gxh, Hohn:2018iwn, Hoehn:2019fsy, Hoehn:2020epv, Castro-Ruiz2020, Hardy:2019cef, yang2020switching, Ballesteros:2020lgl, streiter2020relativistic,Giacomini2020, delaHamette:2020dyi, Krumm:2020fws, tuziemski2020decoherence,  mikusch2021transformation, AhmadAli:2021ajb, Hoehn:2021flk, Giacomini2021, Cepollaro2021, delaHamette:2021oex, delaHamette:2021piz, DelaHamette2021, Castro-Ruiz:2021vnq, Giacomini2022, Kabel2022}).
Such a transformation allows one to associate a reference frame with a  particle that is in a quantum superposition of positions from the perspective of the initial reference frame. 
QRFs have been shown to be relevant for quantum systems in a gravitational field or in a superposition thereof~\cite{Hardy:2019cef, Castro-Ruiz2020,Giacomini2020,Giacomini2021,Cepollaro2021,Giacomini2022,delaHamette:2021oex, DelaHamette2021,Kabel2022, Christodoulou2022a}.

In Section II, our model of the gravitational Aharonov-Bohm experiment was described in a QRF associated with the detector particle D, which we may conceptualize as the ``laboratory'' QRF.  We now use a QRF transformation to describe the same model in a QRF associated with the test particle (particle A).  Specifically, we transform to the QRF centered on the position of particle A.  The position operators of the other particles correspond, in this QRF, to the displacements between those particles and the trajectory of particle A.
This transformation is accomplished by 
\begin{equation}
    \hat{S} = \mathcal{\hat{P}}_{AD} e^{\frac{i}{\hbar}\hat{x}_A \hat{p}_B}. \label{Eq:QRFtranformation}
\end{equation}
Here $\mathcal{\hat{P}}_{AD}$ is the parity-swap operator, defined as
\begin{align}
    \mathcal{\hat{P}}_{AD} \ket{x}_A &= \ket{-x}_D, \\
    \mathcal{\hat{P}}_{AD} \ket{p}_A &= \ket{-p}_D.
\end{align}
The position operators $\hat{y}_B$, $\hat{y}_D$ and momentum operators $\hat{\pi}_B$, $\hat{\pi}_D$ in particle A's QRF are given by 
\begin{equation}
     \begin{split}
         \hat{S}\hat{x}_A\hat{S}^\dagger = -\hat{y}_D, &\qquad 
    \hat{S}\hat{x}_B\hat{S}^\dagger = \hat{y}_B -\hat{y}_D, \\
    \hat{S}\hat{p}_A\hat{S}^\dagger = -\hat{\pi}_B - \hat{\pi}_D, &\qquad
    \hat{S}\hat{p}_B\hat{S}^\dagger = \hat{\pi}_B.
     \end{split}
\end{equation}
The unperturbed trajectories are 
\begin{align}
    &y_{B,i} (t) = x_s(t) - x_i (t), \qquad i=1,2\\
    &y_{D,i} (t) = - x_i (t),
\end{align}
where $x_i(t)$ and $x_s(t)$ were defined respectively in Eq.\,\eqref{eq:trajAfromC} and in Eq.\,\eqref{eq:trajBfromC}.
In the time intervals $(0, T)$ and $(T, 2T)$, the Hamiltonian $H_{BD}^{(A)}$ that includes the gravitational interaction with particle B is given by 
\begin{align} \label{eq:HamA}
    \hat{H}_{BD}^{(A)} &= \hat{S}\hat{H}_{AB}^{(D)}\hat{S}^\dagger + i \hbar \frac{d \hat{S}}{dt} \hat{S}^\dagger \\
    &=\frac{(\hat{\pi}_B + \hat{\pi}_D)^2}{2 m} + \frac{\hat{\pi}^2_B}{2 M} + U(|\hat{y}_B|) - M a (\hat{y}_B - \hat{y}_D).
\end{align}
The definition of the interferometer phase, given in Eq.~\eqref{Eq:intPhase}, is unchanged by the QRF transformation.

\begin{figure*}[t]
\centering
\includegraphics[width=1.0\linewidth]{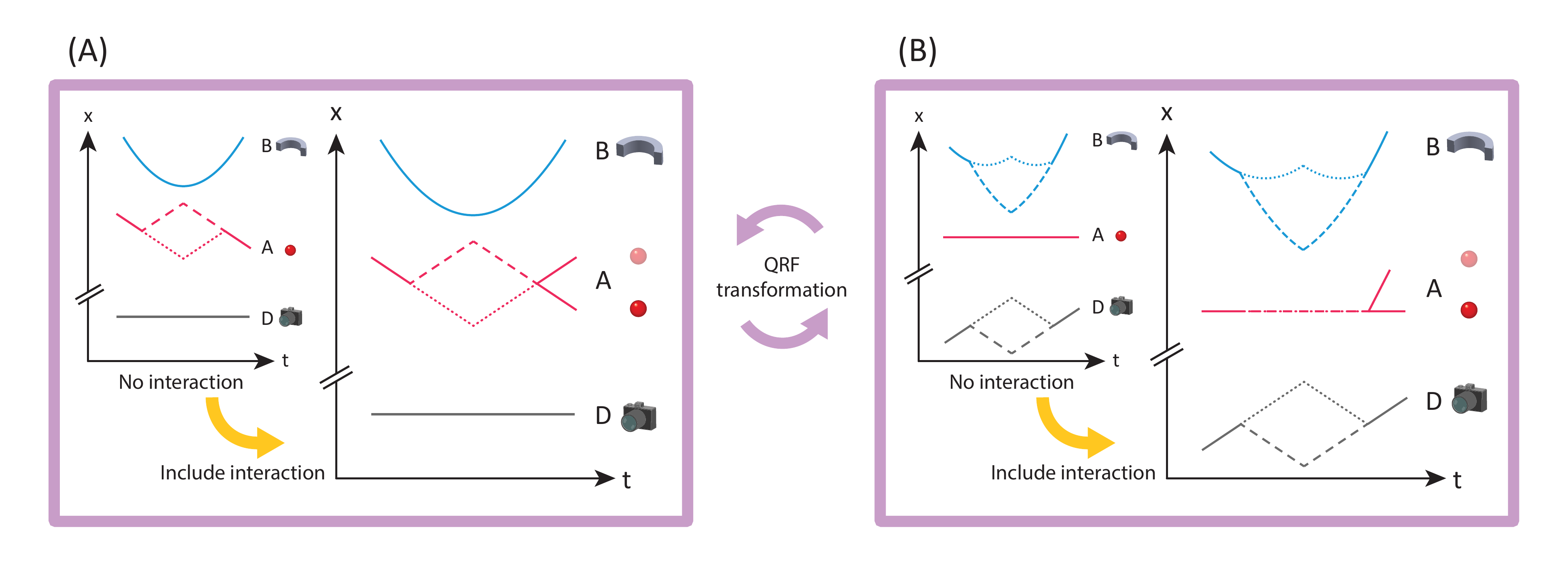}
\caption{\label{Fig:3}(A):  Model of the gravitational Aharonov-Bohm experiment in the QRF of particle D.  The gravitational interaction between test particle A and source particle B induces a phase shift.  (B):  Same experiment in the QRF of unperturbed particle A.  In this QRF, the phase shift of particle A arises because particle A evolves in the gravitational field superposition sourced by particle B.}
\end{figure*}

Fig.~\ref{Fig:3} illustrates the experiment in particle A's reference frame.  When the gravitational interaction is included, the quantum state in the reference frame of particle A takes the form~\cite{Belenchia2018, Danielson2022, Chen2022}
\begin{align} 
    \ket{\psi}^{(A)}(t) = \frac{1}{\sqrt{2}}& \Big(\ket{-x_1}_D \ket{x_s-x_1}_B \ket{\beta_1}_\text{G}  \nonumber \\  &+ e^{i \theta(t)}\ket{-x_2}_D \ket{x_s - x_2}_B \ket{\beta_2}_\text{G} \Big) \label{Eq:quantumStateRefA}  
\end{align}
where the subscripts $B$ and $D$ indicate the states of particles B and D, and $\beta_i$ represents the gravitational field sourced by point particles of mass $m$ and $M$ at positions $0$ and $x_s - x_i$, respectively.  For a gravitational source in the quantum state of Eq.\,\eqref{Eq:quantumStateRefA}, the quantum state of the gravitational field can be derived as explained below Eq.~\eqref{eq:gravitystateD}. We expect the QRF transformation to map Eq.\,\eqref{eq:gravitystateD} into Eq.\,\eqref{Eq:quantumStateRefA} because the gravitational quantum states $\ket{\beta_1}$ and $\ket{\beta_2}$ differ only through the positions of the sources.
 Note that in this reference frame, particles B and D are in an entangled superposition state, and the gravitational field is also in an entangled superposition state with length scale $\hbar k T/m$.

We now need to interpret $\Delta \phi$, the phase shift of particle A due to the gravitational interaction, in the QRF associated with particle A:  
\begin{align}
    \Delta \phi &= \frac{1}{\hbar} \int_{t = 0}^{2T} \left[ U(|y_{B,1}|) - U(|y_{B,2}|) \right] dt  \\
    &= -\frac{G m M}{\hbar} \int_{t = 0}^{2T} \left[ \frac{1}{|x_1 - x_s|} - \frac{1}{|x_2 - x_s|} \right] dt. 
\end{align}
In this case, particle A is located at the origin~\cite{footnote8}, so $\Delta \phi$ cannot be explained in terms of the different paths taken by particle A. Instead, particle B is in a quantum superposition of two semiclassical states, and hence the phase is due to the gravitational field being in a quantum superposition at the location of particle A (see Eq.~\eqref{Eq:quantumStateRefA}). From this perspective, the phase can be obtained as
\begin{equation}
    \phi = \frac{m}{\hbar} \left(\mathcal{V}_1 - \mathcal{V}_2 \right),
\end{equation}
where $\mathcal{V}_i$, $i=1,2$ are the two different gravitational potentials associated with the two different positions of particle B, evaluated at the position of particle A and integrated over the interferometer time. 
Note the distinction between the $\mathcal{V}_i$ in this expression, which indicate two different gravitational potentials, and the $\mathcal{V}(x_i)$ in Eq.~\eqref{Eq:deltaPhiD}, which indicates one gravitational potential evaluated along two different trajectories. 
Quantitatively, this result agrees with the calculation in particle D's reference frame (as expected, since $\Delta \phi$ is an observable). 

Comparing these two descriptions of the system allows us to draw several conclusions.  First, since the reference frames of particles A and D can both be used to calculate the observable $\Delta \phi$, the experiment provides no reason to prefer one reference frame over the other. Second, the expression for $\Delta \phi$ takes the same form in both reference frames.  This occurs because the phase shift depends only on the relative configuration of the masses interacting gravitationally, namely the relative displacement between particles A and B. This is a manifestation of the relational character of general relativity, which has been widely confirmed for classical sources \cite{Will2014}. The invariance of physical laws under coordinate transformations (general covariance) is familiar from general relativity, where the laws of physics are valid regardless of the choice of a specific reference frame (relativity principle).  This experiment exhibits a quantum version of the relativity principle, where the physical laws are valid in any QRF.  

In general, QRF transformations demonstrate that superposition and entanglement are relative properties \cite{Giacomini2019}.  For example, the superposition of particle A in the QRF of particle D is equivalently represented as entanglement between particles B and D in the QRF of particle A.  In the gravitational Aharonov-Bohm experiment, the relative superposition between the test particle and the source mass gives rise to a gravitational field superposition in every QRF.  Note that in previous experiments \cite{Colella1975} where the gravitational field is approximately uniform over the length scale of the apparatus, the QRF transformation in Eq.~\eqref{Eq:QRFtranformation} leaves the gravitational field of the source mass in a state that is not resolvably entangled with the test mass trajectory.  As implied by the equivalence principle, local experiments cannot detect a gravitational field, even if the gravitational field is generated by a superposed source \cite{Bose2022a} (see Appendix 3).

\section{VI.  Comparison to the proposed BMV experiments}

\begin{figure}[h]
\centering
\includegraphics[width=1.0\linewidth]{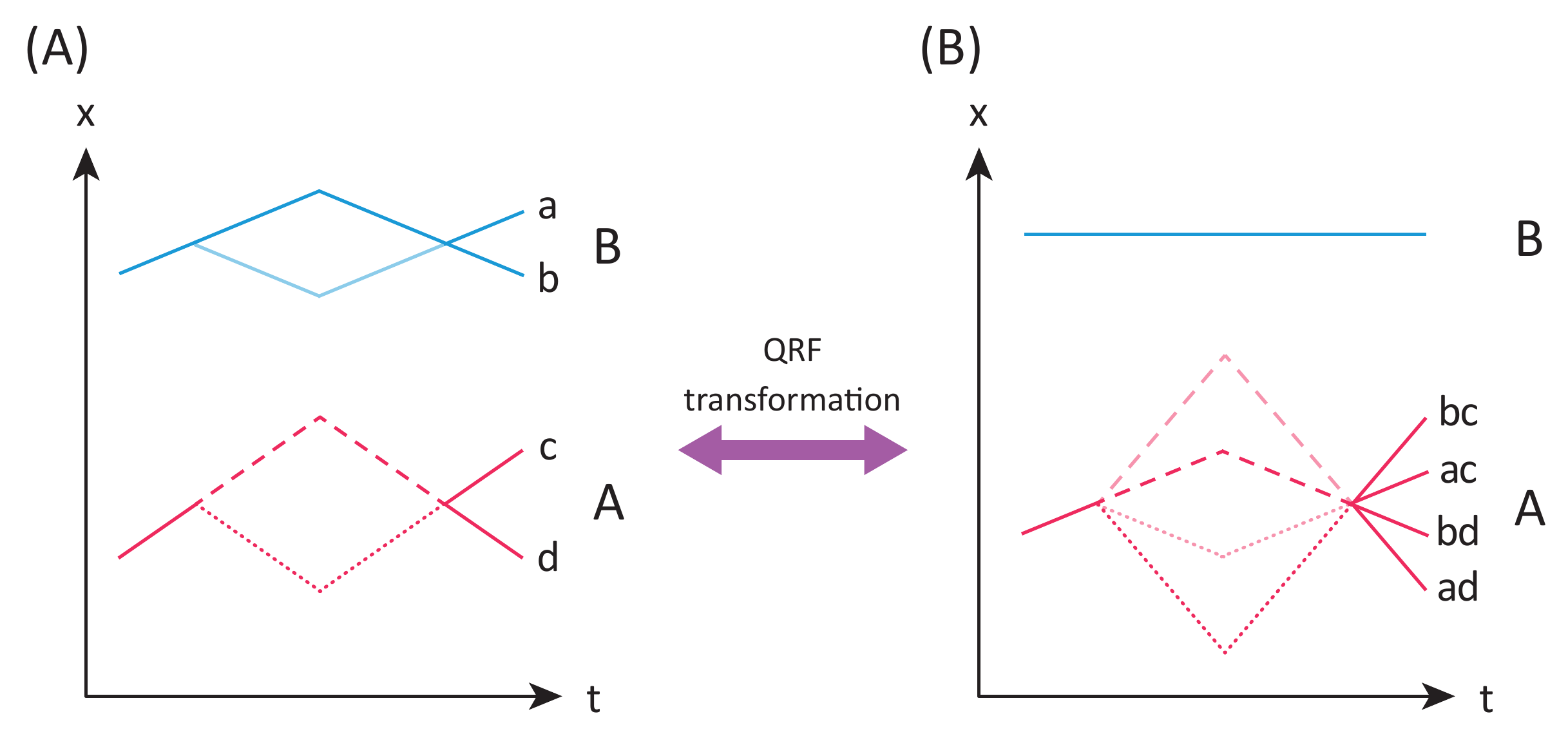}
\label{Fig:4}
\caption{(A):  BMV experiment designed to detect gravitationally mediated entanglement generation between massive particles A and B.  (B):  Same experiment in the QRF of particle B.  If the QRP is valid, each BMV experiment has the same measurement outcome as a corresponding single-particle superposition experiment.}
\end{figure}

\begin{figure*}[t]
\centering
\includegraphics[width=1.0\linewidth]{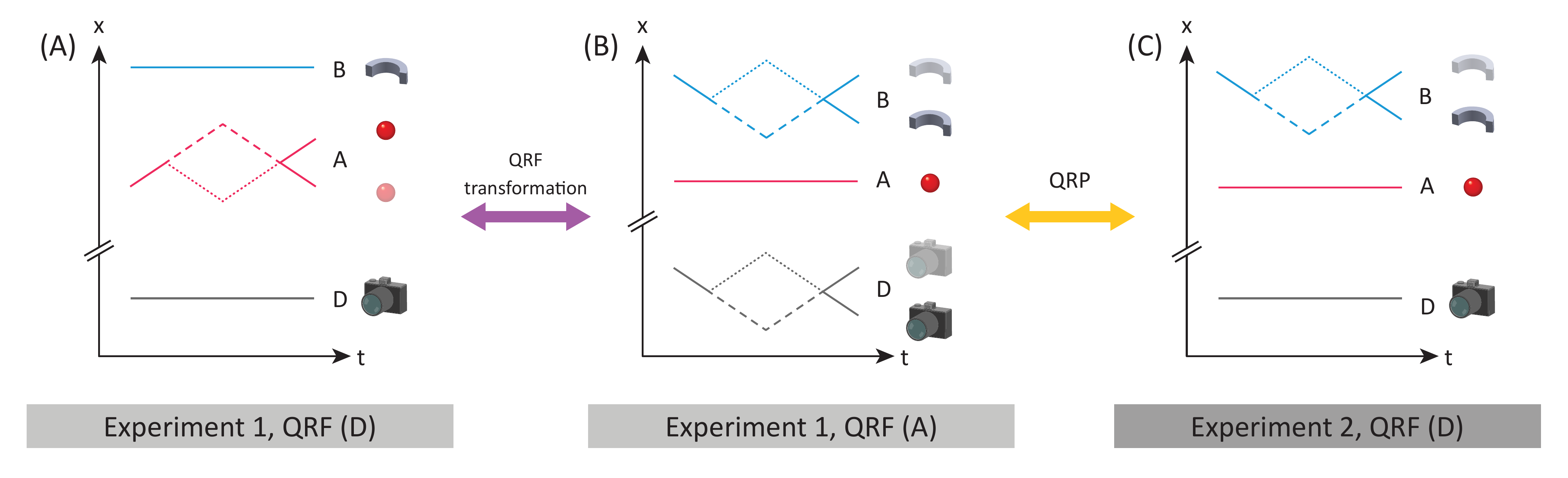}
\label{Fig:5}
\caption{(A):  Simplified model of the gravitational Aharonov-Bohm experiment, represented in the QRF of the detector.  Red lines:  trajectory of test particle A.  Blue line:  trajectory of source mass B, with curvature suppressed for clarity.  Gray line:  trajectory of detector D.  (B):  Gravitational Aharonov-Bohm experiment in the QRF of the test particle.  In this QRF, the test particle is at rest, while the source mass and detector are in an entangled superposition state.  (C): Hypothetical experiment in which the source mass is placed in superposition relative to the test particle and detector. The relative quantum state of the test particle and source mass is the same in both experiments.  By the quantum relativity principle, Experiment 1 has the same measurement outcome as Experiment 2.  Performing an experiment like the one depicted in (C) would test the QRP.}
\end{figure*}

By assuming that the quantum relativity principle is valid, the results of future experiments to detect gravitationally mediated entanglement generation \cite{Bose2017,Marletto2017} can be inferred from gravitational Aharonov-Bohm-like experiments.  This follows from the fact that QRF transformations can map entangled states into single-particle superposed states and vice versa.  Hence, BMV experiments can be related to measurements of single-particle superpositions by QRF transformations.  Fig. 4A illustrates a generic BMV experiment, where two massive particles become entangled due to their gravitational interaction.  Fig. 4B shows the same experiment in the QRF of one of the particles.  In this QRF, one of the massive particles is in a well-localized state, and the other is in a superposition state.  The signature of entanglement generation (namely, the probabilities of finding the system in each of its four output states ac, ad, bc, and bd) appears identically in both QRFs.

According to the QRP, the experiment in Fig. 4A is equivalent to the experiment in Fig. 4B performed in the laboratory reference frame.  Along with the trivial detection of particle B's location, the measurement of particle A encodes the outcome of the BMV experiment in the probabilities of observing particle A in each of the four output ports.  In other words, if the QRP is valid, then the result of a measurement like the one in Ref. \cite{Overstreet2022} (but with the test particle superposed into four arms, rather than two) determines whether or not a BMV experiment would detect entanglement.  

The QRP thus represents the key assumption that links the gravitational Aharonov-Bohm experiment and proposed BMV experiments.  The additional value of performing a BMV experiment is that it would provide the first evidence for or against the QRP in the gravitational sector.  Alternatively, the QRP could be tested (and semiclassical gravitational theories constrained) by means of single-particle superposition experiments.  Fig. 5 illustrates the type of measurement that needs to be performed to validate the QRP:  a detection of the gravitational influence of a localized lower-mass particle on a higher-mass particle in a superposition state.  Since the QRP can be tested with measurements on single particles, the direct observation of gravitationally mediated entanglement generation may ultimately not be necessary to constrain semiclassical gravity.  The key experimental challenge is simply to create a spatial superposition of a massive particle and measure its gravitational interactions with sufficient precision.  Like the proposed BMV experiments, such a measurement is not technologically feasible at present.    

\section{VII.  Results and discussion}

In the previous sections, we gave a quantum-mechanical description of the gravitational Aharonov-Bohm experiment that is compatible with three general principles: (1) the existence of gravitational fields, (2) the field energy principle, and (3) the quantum relativity principle.  Here, we argue that the validity of these principles, combined with the current experimental observations, implies that the gravitational field is in a quantum superposition state.  In other words, if we assume that certain subsets of these principles are valid, then this experiment demonstrates a gravitational superposition state even in the context of potential alternatives to quantum mechanics.  

First, assume that (1) and (2) are valid.  For a contradiction, suppose that a theory describes the gravitational field in Ref. \cite{Overstreet2022} as a classical state.  Then in order to be consistent with classical mechanics, the gravitational field energy must be a function of the test particle's probability distribution.  On the other hand, in order to be consistent with the experimental results in Section 4 of this work, the theory must predict that the interferometer's phase shift is independent of the test particle's probability distribution.  The phase shift is thus independent of the gravitational field energy, and the theory violates (2).  
    
The same conclusion holds if we instead assume that (1) and (3) are valid.  For each of the gravitating objects (test mass and source mass), there exists a quantum reference frame in which the object is well-localized at some position $\mathbf{x_i}$.  In that reference frame, by (1) and (3), the object sources a gravitational field $\mathbf{g_i}(\mathbf{x} - \mathbf{x_i})$.  We can then write the full gravitational field in a common reference frame by means of quantum reference frame transformations.  Since there is a relative superposition between the test mass and source mass, there is a relative superposition between their gravitational fields as well, i.e. the full gravitational field is in a superposition state.  Moreover, the gravitational field superposition has an observable consequence in the experiment because gravity induces the phase shift in the interferometer, so the experiment demonstrates the existence of the superposition state. 

To deny that the gravitational field is in a superposition state in the experiment of Ref. \cite{Overstreet2022}, theories of gravity in which (1) is valid must reject both (2) and (3).  As an example, consider the Schr\"{o}dinger-Newton theory \cite{Carlip2008}, which treats the gravitational field as a fundamentally classical entity that is sourced by the probability distribution of matter.  The equation of motion for this theory contains a gravitational self-interaction term that prevents the creation of large spatial superpositions of massive particles, where the critical length scale depends sensitively on the mass.  In the laboratory reference frame, the Schr\"{o}dinger-Newton equation predicts that the gravitational field during the interferometer is in a classical state of the form $|A_1|^2 \ket{\alpha_1}_\text{G} \bra{\alpha_1}_\text{G} + |A_2|^2\ket{\alpha_2}_\text{G} \bra{\alpha_2}_\text{G}$, where $A_i$ is the wavefunction amplitude of the test mass on trajectory $i$ \cite{footnote13}.  The energy of the gravitational field is therefore 
\begin{equation}
    E^\text{S-N} = |A_1|^2 E(\alpha_1) + |A_2|^2 E(\alpha_2).
\end{equation}
The predicted (and observed) independence of the interferometer phase shift from $A_1$ and $A_2$ indicates that the phase shift is not a function of $E^\text{S-N}$, violating (2).  Furthermore, the Schr\"{o}dinger-Newton theory implicitly contains a preferred quantum reference frame.  Although the Schr\"{o}dinger-Newton equation evaluated in the laboratory frame appears to be consistent with the experimental results of Ref.~\cite{Overstreet2022}, in the quantum reference frame of the test mass, the Schr\"{o}dinger-Newton equation inconsistently predicts the collapse of the source mass superposition.  The theory thus violates (3) \cite{footnote7}.  A similar consideration can be also made in the case of Penrose's spontaneous state reduction~\cite{Penrose2014,Giacomini2022}, where the equivalence of all quantum reference frames can be stated in terms of a generalisation of the Einstein Equivalence Principle~\cite{Giacomini2020}.

In conclusion, if one makes certain general assumptions about the existence and properties of gravitational fields, then the existence of a gravitational superposition state has already been demonstrated experimentally.  Future experiments will seek to obviate the need for these assumptions, providing even more stringent tests of gravity's fundamental nature.  Either the observation of gravitational entanglement generation in a BMV experiment, or a sufficiently precise gravitational measurement using a superposed massive particle, would demonstrate the existence of a nonclassical state of the gravitational field even without assuming principle (2) or (3).

\vspace{0.05in}
  
 \noindent\textbf{Acknowledgments.} F.G. would like to thank Markus Aspelmeyer and Carlo Cepollaro for useful discussions. C.O. acknowledges support from the Q-FARM Bloch Fellowship.  F.G. acknowledges support from the Swiss National Science Foundation via the Ambizione Grant PZ00P2-208885, from the ETH Zurich Quantum Center, from the John Templeton Foundation, as part of the \href{https://www.templeton.org/grant/the-quantuminformation-structure-ofspacetime-qiss-second-phase}{‘The Quantum Information Structure of Spacetime, Second Phase (QISS 2)’ Project}, and from Perimeter Institute for Theoretical Physics. Research at Perimeter Institute is supported in part by the Government of Canada through the Department of Innovation, Science and Economic Development and by the Province of Ontario through the Ministry of Colleges and Universities. J.C. acknowledges support from the National Defense Science and Engineering Graduate Fellowship.  M.K. acknowledges support from the Kwanjeong Educational Foundation.  The work of C.O., J.C., and M.K. was partially supported by the U.S. Department of Energy, Office of Science, National Quantum Information Science Research Centers, Superconducting Quantum Materials and Systems Center (SQMS) under contract number DE-AC02-07CH11359.

\section{Appendix 1:  Relational, reciprocal character of the phase shift}

In this section, we show that the electromagnetic and the linearized gravitational actions are symmetric under exchange of two arbitrary interacting particles. We also show that the actions depend only on the relative position of the source particles. This argument is not intended to be a description of the experimental apparatus in the main text, but rather a more general justification for the quantum relativity principle. 

Let us first review the electromagnetic case. The free action of the electromagnetic field is
\begin{align}
    \mathcal{S}_\text{free} = -\frac{1}{4}\int d^4 x F_{\mu\nu}(x)F^{\mu\nu}(x), \nonumber \\ \qquad F_{\mu\nu}(x) = \partial_\mu A_\nu (x) - \partial_\nu A_\mu (x)
\end{align}
and $A_\mu (x)$ is the vector potential. In addition to the free action, in the presence of a source term there is the matter-field interaction, which can be written as 
\begin{equation}
    \mathcal{S}_\text{int} = \int d^4 x A_\mu(x) J^\mu (x).
\end{equation}
Here, $J^\mu (x)$ is the current density associated with the source. In the presence of two sources with total current $J^\mu (x)= J^\mu_1 (x) + J^\mu_2 (x)$, the principle of linear superposition holds, hence $A^\mu (x)= A^\mu_1 (x) + A^\mu_2 (x)$, where each $A^\mu_i$ with $i=1,2$ solves the wave equation for the corresponding current $J^\mu_i (x)$.

In perturbation theory, the interferometer phase shift can be calculated from the interaction term $S_\text{int}$, which is completely symmetric under exchange of the two sources 1 and 2. This can be easily seen by making use of path-integral methods in quantum field theory~\cite{Greiner:1996zu} and expressing the phase in terms of the Feynman propagator $G^{\rho}_{\nu} (x-y)$ solving the differential equation $\mathcal{K}^{\mu}_{\rho} G^{\rho}_{\nu} (x-y) = \delta^\mu_\nu \delta^{(4)}(x-y)$. Here, $\mathcal{K}^{\mu}_{\rho}$ is a differential operator corresponding to the wave equation, which is obtained via the equations of motion of the free theory. A particular solution of the wave equation is obtained as $A^\mu (x) = \int d^4 y G^{\mu}_{\nu} (x-y)J^\nu (y)$. The Feynman propagator defined in path-integral methods for quantum field theory can be expressed by considering a sum of the retarded and advanced solutions of the Green function (see again~\cite{Greiner:1996zu} for details of the calculation). Removing the self-interaction, we can rewrite
\begin{equation}
    \mathcal{S}_\text{int} = \int d^4 x d^4 y J^\mu_1 (x) G_{\mu\nu} (x-y) J^\nu_2 (y).
\end{equation}
Notice that this expression is valid in general, regardless of the relative magnitude of the charges sourcing the field. This shows that the phase can be equivalently calculated by considering the field sourced by system 1 coupled to the source term of system 2, or vice versa.

The same reasoning applies to the weak gravitational field. From the weak gravity action (see, e.g.\,Ref.~\cite{Maggiore2007})
\begin{align}
    \mathcal{S}_\text{G} = \frac{1}{4\kappa}& \int  d^4x  \big(-	\partial_{\mu} {h}_{\alpha \beta} \partial^{\mu} {h}^{\alpha \beta} + 	\partial_\mu {h}  \partial^\mu {h} \nonumber \\	&-2 \partial_{\mu} {h}^{\mu \nu} \partial_\nu {h} +2 \partial_\alpha {h}_{\mu \nu} \partial^\mu {h}^{\alpha\nu} \big) \nonumber \\ &+ \frac{1}{2}\int d^4x\, {h}_{\mu \nu} T^{\mu \nu}, 
\end{align}
where $T^{\mu\nu}$ is the stress-energy tensor, it is possible to cast the equations of motion in the form
\begin{equation}\label{eq:EOM_weakgravity}
    \mathcal{K}_{\alpha\beta}^{\mu\nu} h^{\alpha\beta}(x) = T^{\mu\nu}(x).
\end{equation}
Explicitly, one finds
\begin{align}
    \mathcal{K}_{\alpha\beta}^{\mu\nu} = &\delta^\mu_{\lbrace \alpha}\delta^\nu_{\beta\rbrace}\Box -\eta^{\mu\nu}\eta_{\alpha\beta}\Box + \eta^{\mu\nu}\partial_\alpha\partial_\beta \nonumber \\ &+ \eta_{\alpha\beta}\partial^\mu\partial^\nu - \delta^{\lbrace \mu}_{\lbrace \alpha}\partial^{\nu\rbrace}_{\phantom\lbrace} \partial^{}_{\beta\rbrace},
\end{align}
where $\Box = \partial_\rho \partial^\rho$ and the curly brackets indicate the symmetrization of the indices. A particular solution of Eq.~\eqref{eq:EOM_weakgravity} is
\begin{equation}
    h^{\alpha\beta}(x) = \int d^4 y G^{\alpha\beta}_{\mu\nu}(x-y) T^{\mu\nu}(y),
\end{equation}
where $\mathcal{K}_{\alpha\beta}^{\mu\nu} G^{\alpha\beta}_{\rho\sigma}(x-y) = \delta^\mu_{\lbrace \rho} \delta^\nu_{\sigma \rbrace} \delta^{(4)}(x-y)$. In the linearized regime of gravity, the principle of linear superposition holds, so that in the presence of two sources of the gravitational field $T^{\mu\nu}(x) = T^{\mu\nu}_1(x)+ T^{\mu\nu}_2(x)$ the full metric is the sum of the two independent solutions $h^{\mu\nu}_{1}(x)$ and $h^{\mu\nu}_{2}(x)$ corresponding respectively to the stress energy tensors $T^{\mu\nu}_1(x)$ and $T^{\mu\nu}_2(x)$. We then remove the self-interaction analogously to the electromagnetic case, and we obtain 
\begin{equation} \label{eq:Gactionint}
    \mathcal{S}_\text{G,int} = \frac{1}{2}\int d^4x d^4 y\, T_{\alpha\beta}^1(x) G^{\alpha\beta}_{\mu\nu}(x-y) T^{\mu \nu}_2(y).
\end{equation}

From the symmetry of the action, it is now clear that it is equivalent to consider the gravitational field sourced by system 1, namely $h^{\mu\nu}_1 (x)= \int d^4 y\, G_{\alpha\beta}^{\mu\nu}(x-y) T^{\alpha\beta}_1(y)$, coupled to system 2 via its stress-energy tensor $T^{\mu\nu}_2(x)$ or, vice versa, the gravitational field sourced by system 2, namely $h^{\mu\nu}_2 (x)= \int d^4 y \, G_{\alpha\beta}^{\mu\nu}(x-y) T^{\alpha\beta}_2(y)$, coupled to system 1 via $T^{\mu\nu}_1(x)$. Notice that the propagator $G_{\alpha\beta}^{\mu\nu}(x-y)$ is symmetric in its indices and is the same in both cases, as it is determined by the free theory.  Furthermore, the propagator depends only on the position difference between systems 1 and 2.  

For two localized particles, we can choose a coordinate system that is centered on one of the particles, say in an initial position $x_0$. It is easy to see that this choice of coordinates leaves the action invariant by changing the coordinates to $x'= x+x_0$, $y' = y + x_0$ in Eq.~\eqref{eq:Gactionint}.  Then, when one of the particles is in a quantum superposition of two states that are well-localized in the position basis, it can be shown~\cite{Chen2022} that the gravitational field can be obtained (in the linearized regime) as the quantum superposition of the ground state configurations of the field associated with each state. Thanks to the fact that the action of Eq.~\eqref{eq:Gactionint} is invariant for the single transformation, we conclude that the difference of the action along the two paths is also invariant under a change of QRF.
In the case of the gravitational Aharonov-Bohm experiment~\cite{Overstreet2022}, which can be described in the Newtonian limit, this means that it is equivalent to use the laboratory description in which particle B is localized while particle A is in a quantum superposition of trajectories, or the QRF associated with the position of particle A, in which particle B is in a quantum superposition of trajectories \cite{footnote10}.

\section{Appendix 2: Example semiclassical gravitational model} 

As discussed in Section VII, the results presented in this work do not constrain the Schr\"{o}dinger-Newton theory or other semiclassical models in which each quantum trajectory is coupled locally to the gravitational field. 
 However, our measurements do constrain models in which a quantum particle is coupled to the gravitational field in some other way, e.g. through the expectation value of its position.  For example, we consider a semiclassical gravitational model with the following properties:  

(i)  Gravity is sourced by the expectation value of each quantum particle's position, and

(ii)  Quantum particles couple to gravity through the expectation value of position only.  That is, the expectation value of a quantum particle's position is influenced by the gravitational field, but the higher moments of its wave function are not affected.  

Specifically, we consider the gravitational interaction between two particles A and B of mass $m$ and $M$, respectively, in the limit where {$m \ll M$}.  We assume that particle B is well-localized on trajectory $x_s(t)$, and we neglect the gravitational deflection of particle B due to particle A.   

The gravitational field in this model is given by 
\begin{equation}
    g(x, t) = - \frac{G M}{|x - x_s(t)|^2} \hat{r}_s - \frac{G m}{|x - x_{CM}(t)|^2} \hat{r}_{CM}, \label{eq:SemiclassicalFieldGeneral2}
\end{equation}
where $x_{CM}$ is the expectation value of particle A's position, $\hat{r}_s$ is a unit vector parallel to $x - x_s$, and $\hat{r}_{CM}$ is a unit vector parallel to $x - x_{CM}$.  For an interferometer configuration in which particle A is in a quantum superposition of two orthogonal states corresponding to two trajectories $x_1(t)$ and $x_2(t)$ with probability densities $P_1$ and $P_2$, respectively, we have $x_{CM} = P_1 x_1 + P_2 x_2$.  Note that the gravitational field remains classical even though particle A is in a superposition state.  

The gravitational coupling of particle A in this model is motivated by the field energy principle.  From Eq.~\eqref{eq:SemiclassicalFieldGeneral2}, the gravitational interaction energy $E$ is given by
\begin{equation}
    E = - \frac{G M m}{|x_{CM} - x_s|},
\end{equation}
which suggests that there should be a gravitational force influencing the time evolution of $x_{CM}$.
In our model, this force is exerted uniformly across the wave function of particle A.  Thus, the trajectories $x_1$, $x_2$, and $x_{CM}$ are solutions of the equation of motion 
\begin{equation}
    \ddot{x}_i = - \frac{G M}{r^2} \hat{r}. \label{eq:SemiclassicalEqnofMotion2}
\end{equation}
for $i \in \{1, 2, CM\}$, where $r \equiv x_{CM} - x_s$ and $\hat{r}$ is a unit vector parallel to $r$.

In the classical limit where the wave packet separation $|x_1 - x_2|$ goes to zero, we have $x_1 = x_2 = x_{CM}$, and Eq.~\eqref{eq:SemiclassicalEqnofMotion2} reduces to the classical equation of motion.  Thus, classical gravitational experiments are consistent with this model.  Furthermore, Eq.~\eqref{eq:SemiclassicalEqnofMotion2} correctly predicts the result of the COW experiment \cite{Colella1975}, where the gravitational field is approximately uniform at the length scale of the wave packet separation.  

However, this model is inconsistent with the results of Ref. \cite{Overstreet2022} and the data in Fig. 2 of this work.  Specifically, Eq.~\eqref{eq:SemiclassicalEqnofMotion2} implies that the midpoint deflection (and hence the phase shift) of the $52\hbar k$ interferometer is rescaled by changing the probability distribution between the two arms.  For the experimental configuration shown in Fig. 2, a numerical simulation of this semiclassical model predicts a phase shift of \{-0.198, -0.374, -0.394\} for probability density \{0.25, 0.5, 0.75\} on the upper interferometer arm.  These predictions are inconsistent with the phase shift data \{-0.23, -0.24, -0.27\} shown in Fig. 2.  

\section{Appendix 3:  The equivalence principle for quantum sources}

According to the equivalence principle, gravity cannot be observed locally \cite{footnote9}.  As applied to classical gravitational sources, the equivalence principle has been tested to high precision with classical \cite{Toboul2017} and quantum \cite{Asenbaum2020a} test masses.  The QRP motivates an extension of the equivalence principle to the case where a gravitational source is in a quantum superposition state~\cite{Giacomini2020,Cepollaro2021} (see also~\cite{hardy2020implementation}).  We may state the \textit{equivalence principle for quantum sources} as follows:  no local experiment can detect the gravitational field produced by a superposed gravitational source.  

\begin{figure}[h]
\centering
\includegraphics[width=1.0\linewidth]{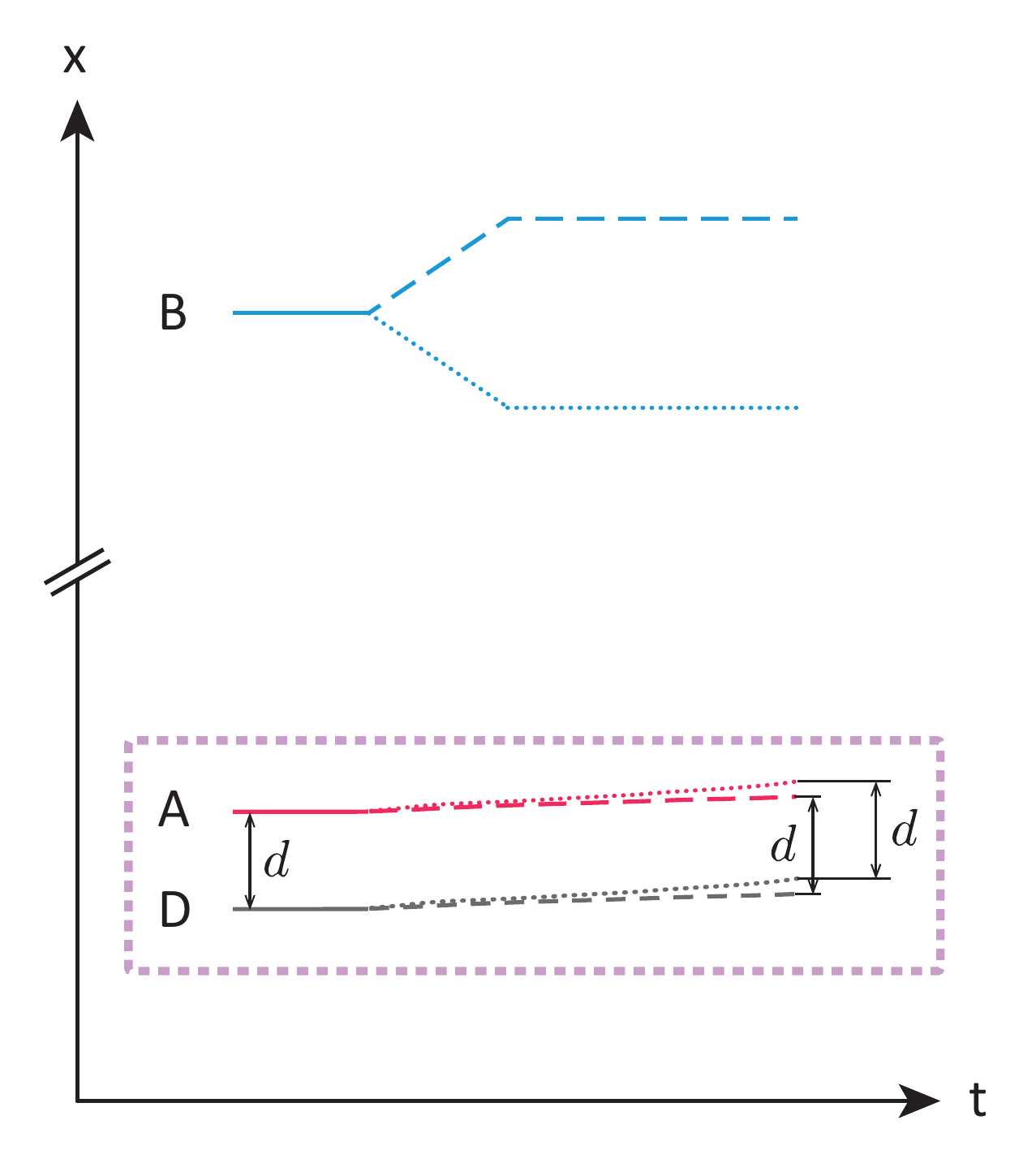}
\label{Fig:app2}
\caption{Accelerometer (test particle A and reference D) in the gravitational field of a quantum source (B).  The distance $d$ between A and D is small enough that the accelerometer can be considered a local system.  According to the equivalence principle, even though the trajectories of A and D become entangled with the trajectories of B, the distance between A and D is unaffected.  Thus, the accelerometer cannot detect the gravitational field from B.}
\end{figure}

Fig. 6 depicts a thought experiment that attempts to observe the gravitational field from a quantum source by using an accelerometer.  The accelerometer consists of a test particle (which may be classical or quantum) and a reference; the accelerometer measures the relative acceleration between the test particle and the reference.  As the system evolves in time, the gravitational field entangles the trajectories of the test particle and the reference with the trajectory of the gravitational source.  However, as long as the distance between the test particle and reference is sufficiently small that gravitational tidal effects can be ignored, there is no relative acceleration between the test particle and the reference on either branch of the accelerometer's wave function.  This follows from the equivalence principle, which states that the test particle and reference fall identically in a gravitational field, regardless of their masses or compositions.  

As long as the accelerometer's time evolution is linear, no information can be transmitted from one branch of its wave function to the other, and therefore the accelerometer cannot detect the gravitational field from the quantum source.  If particles A and D are replaced by clocks, an analogous argument shows that there is no observable redshift due to particle B (see Ref.~\cite{Cepollaro2021} for an explicit derivation of this result in quantum reference frames). The equivalence principle for quantum sources is thus satisfied in standard quantum mechanics.  Furthermore, any theory of gravity that incorporates the (classical) equivalence principle as well as linear time evolution also includes the equivalence principle for quantum sources.

\section{Appendix 4:  Entanglement in the gravitational Aharonov-Bohm experiment}

\begin{figure}[h]
\centering
\includegraphics[width=1.0\linewidth]{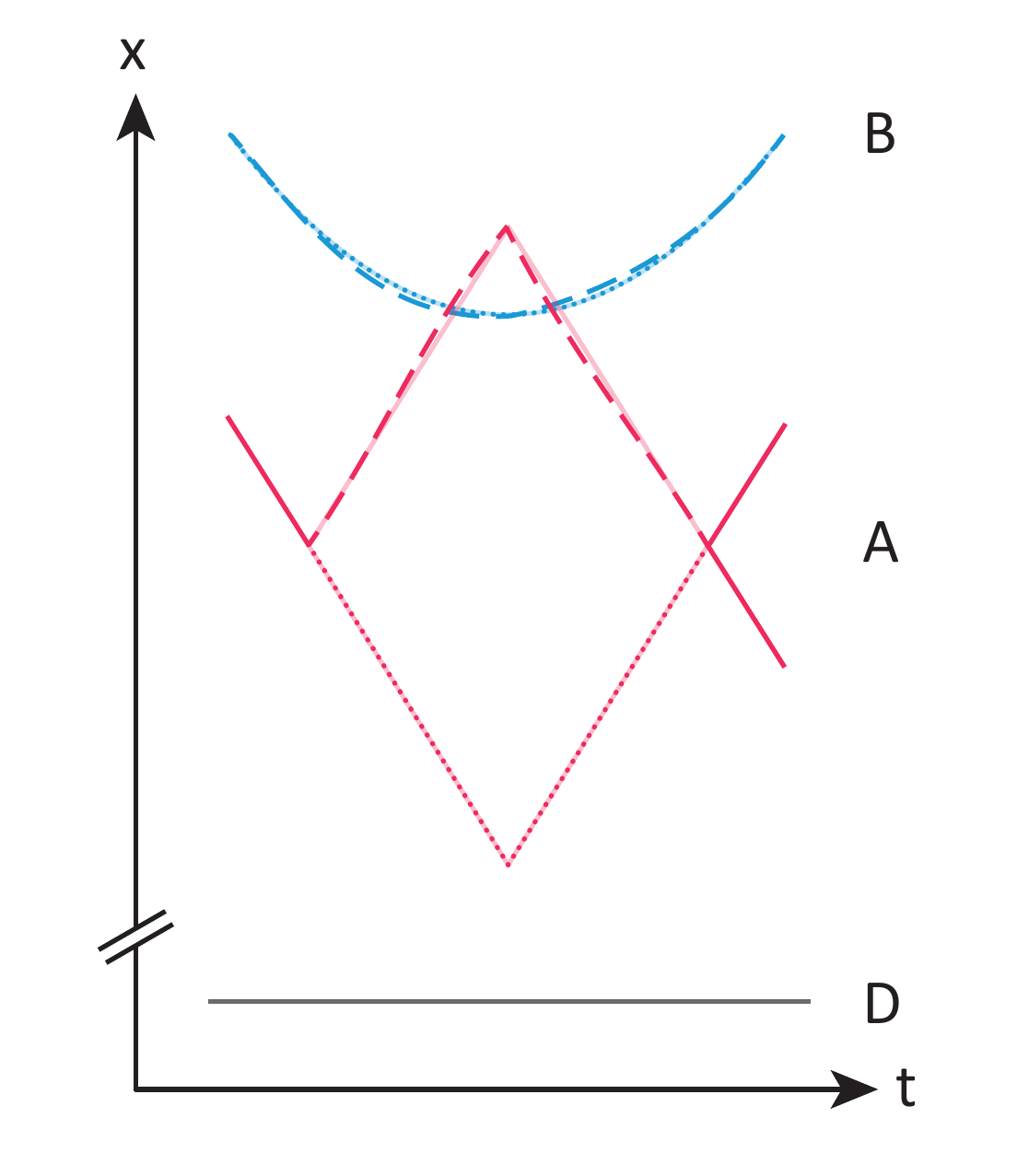}
\label{Fig:app3}
\caption{Simplified model of the gravitational Aharonov-Bohm experiment with effects on source mass included. Solid curves:  unperturbed trajectories.  Dotted and dashed curves:  trajectories including gravitational back action.  The dashed (dotted) trajectory of particle A is entangled with the dashed (dotted) trajectory of particle B.  Diagram is not to scale.  In Ref. \cite{Overstreet2022}, the gravitational deflection of particle A is independently measured.  Owing to the quantum uncertainty in particle B's position, the back action on particle B's trajectory cannot be observed.}
\end{figure}

Since BMV experiments are usually discussed in terms of gravitationally mediated entanglement generation, one might wonder how the BMV result can be inferred from the gravitational Aharonov-Bohm experiment without mentioning entanglement.  In fact, if the QRP is valid, there is gravitational back action and entanglement generation in the experiment of Ref. \cite{Overstreet2022} (see Fig. 7).  Specifically, the gravitational interaction between the source mass and the upper interferometer arm entangles their trajectories, causing phase shifts and deflections of the source mass trajectories during the interferometer.  As pointed out by Vaidman in the context of the electromagnetic Aharonov-Bohm effect, the phase shift of the test particle due to the interaction is equivalent to the phase shift of the source \cite{Vaidman2012}.  At a particular launch height, the net deflections of the source mass and test mass vanish at the time of each beamsplitter pulse.

It is not possible to directly observe the back action on the source mass during the interferometer, even in principle, because the trajectory perturbation due to the back action is much smaller than the quantum uncertainty in the source mass position.  For instance, even if the quantum uncertainty in the velocity of the source mass in Ref. \cite{Overstreet2022} was as large as $1$ mm/s, the quantum uncertainty in its position would still be at least $10^{-32}$ m, which is larger than the maximum gravitational deflection of the source mass.  Nevertheless, assuming that the QRP is valid, a self-consistent description of the experiment requires the entanglement generation and associated back action to be included.

%Bibliography
% \bibliographystyle{spphys}
% \bibliography{library}

\begin{thebibliography}{10}
\bibitem{footnote4}
Since the center-of-mass trajectories of the wave packet on each interferometer arm are sufficient to compute the interferometer phase, we will not model the diffraction of the wave packet.
\bibitem{footnote11}
The trajectory perturbations of particle A are of order $10$~nm or smaller, compared to the $\sim 10$~cm curvature scale of $U$.
\bibitem{footnote8}
As in Section 2, we can ignore the trajectory perturbations of particle A for the purpose of this perturbative phase shift calculation.
\bibitem{footnote13}
Note that maximally mixed states correspond to classical states.
\bibitem{footnote10}
In the QRF of particle A, in order for the gravitational field states associated with the two trajectories of particle B to be orthogonal, it is necessary for the distance between the two trajectories to be larger than the localization of particle B on each trajectory. Equivalently, in the laboratory frame, the center of mass of particle B must be localized to an accuracy $\delta l \ll L$, where $L$ is the spatial separation between the two trajectories of particle A.  This condition is satisfied experimentally.
\bibitem{footnote9}
Here we use ``local'' in the general-relativistic sense to indicate a system that is sufficiently small that spacetime curvature (e.g. gravity gradients) can be neglected within the system.
\end{thebibliography}

%Manual bibliography

\end{document}